\documentclass{article}
\usepackage[a4paper, total={6in, 8in}]{geometry}
\usepackage[utf8]{inputenc}
\usepackage{amstext}
\usepackage{graphicx,caption,epstopdf}
\usepackage{gensymb}
\usepackage[caption=false,font=normalsize]{subfig}
\usepackage{xcolor}
\usepackage{cite}
\usepackage{setspace}
\title{Bipartite Dielectric Huygens' Metasurface for Anomalous Refraction}
\author{Abhishek Sharma*, Alex M. H. Wong**}
\date{Department of Electrical Engineering\\
State Key Laboratory of Terahertz and Millimeter Waves\\
City University of Hong Kong, Hong Kong, China.\\
Email Address: $^{*}$abhisheksharma.rf@gmail.com, $^{**}$alex.mh.wong@cityu.edu.hk}

\begin{document}

\maketitle
\begin{abstract}
\emph{Huygens' metasurfaces} -- fundamentally based on \emph{Schelkunoff’s equivalence principle}, Huygens' metasurfaces consist of a two-dimensional array of Huygens' sources formed by co-located orthogonal electric and magnetic dipoles. Such metasurfaces provide electric and magnetic responses to an incoming electromagnetic (EM) wave, leading to unidirectional scattering and $2\pi$ phase coverage. We herein report a \emph{near-reflectionless coarsely discretized dielectric Huygens' metasurface} that performs anomalous refraction, offering a low-loss platform for wave manipulation at high frequencies as compared to their lossy metallic analogue. The coarse discretization dramatically simplifies the design, resulting in a metasurface that is highly efficient, cost-effective and robust. In this paper, the proposed metasurface comprises two meta-atoms per period, and is hence named the \emph{bipartite dielectric Huygens' metasurface}. Through full-wave simulations at 28 GHz, we show that the proposed metasurface can reroute an incident EM wave from $\theta_i=15^{\circ}$ to $\theta_t=-44.5^{\circ}$ with a very high efficiency: 87\% of the scattered power is anomalously transmitted to $\theta_t$. Based on our observations, a coarsely discretized dielectric Huygens' metasurface platform can be efficacious to design meta-devices with multifaceted functionalities in different frequency regimes.
\end{abstract}
\section{Introduction}
Electromagnetic (EM) metasurfaces -- the 2D counterpart of bulk metamaterials, have revolutionized the field of \emph{surface electromagnetics} and opened up a new avenue for exotic EM wave manipulation in a nearly arbitrary manner~\cite{oscar2019,ss2019,cwq2021,as2021,yr2022}. In recent years, \emph{Huygens' metasurfaces} (HMSs) have offered a promising approach for exquisitely controlling the EM waves in a reflectionless manner~\cite{cp2013,ms2013,mchen2018,vga2022}. In general, HMSs are made up of co-located orthogonal electric and magnetic dipoles known as \emph{Huygens' sources}, which manipulate wavefronts utilizing \emph{Schelkunoff's equivalence principle}~\cite{roger2001} -- a rigorous formulation of the classical \emph{Huygens' principle}. Such metasurfaces provide electric and magnetic responses to an incoming EM wave, imparting high transmission and full $2\pi$ phase coverage. HMSs have been widely used to implement a variety of efficient wave manipulating devices, including anomalous reflectors and refractors~\cite{mc2018,amhw2018,ajo2018,cl2019,chen2020}, metalenses~\cite{mcap2019,aeo2020,ql2021}, and specialized antennas~\cite{ap2016,kao2020,as2022}, to name a few. The two well-known unit-cell topologies for realizing HMSs with metallic scatterers (meta-atoms) are: (i) wire-loop meta-atom, which consists of a conductive loaded dipole to control electric response, and a conductive loaded loop for controlling magnetic response~\cite{cp2013,chen2020}, and (ii) stacked layer topology, comprising three to four cascaded impedance sheets, which essentially are bianisotropic structures, providing both equivalent electric and magnetic responses~\cite{mc2018,mcap2019,luke2021}. Other topologies include multi-layered metallic patterns connected through vias~\cite{jpsw2014,fsc2018}, gap surface plasmon resonator~\cite{mk2014}, among many others.

The realization of HMS having metallic inclusions is quite challenging, especially at higher frequencies, such as millimeter-wave (mm-wave), terahertz, and beyond. These challenges are multi-fold. First, the implementation requires vias or stacked-layer topology, leading to fabrication complexities as we go higher and higher in frequency. Second, the metallic scatterers' inevitable ohmic losses degrade metasurface performance, preventing them from being used at high frequencies. Third, the bonding layer used to glue different substrate layers may incur additional losses. 

On the contrary, all-dielectric metasurfaces -- comprising low-loss and high-permittivity dielectric meta-atoms, have been suggested as a low-loss alternative to their lossy metallic counterparts for remarkable EM wave manipulation~\cite{aco2019,is2019,kk2021}. Apart from the low-loss implementation, such metasurfaces offer new opportunities based on the interplay between the electric and magnetic eigenmodes~\cite{kk2021,md2015,yc2018,ard2019,wen2020,liu2020}, which is the subject matter of this paper. \emph{Dielectric Huygens' metasurfaces} (DHMSs) -- a subset of all-dielectric metasurfaces comprise a 2D array of Huygens' sources formed by spectral overlapping of crossed electric and magnetic dipoles or quadrupoles in a single dielectric block~\cite{md2015,liu2020}. The superposition between the electric and magnetic eigenmodes having comparable strength can result in enhanced forward scattering, and suppression of backward scattering similar to the \emph{Kerkers' first condition}~\cite{mk1983}. Such spectral overlapping can be achieved by appropriately tuning the physical parameters of the dielectric resonator (DR), leading to full $2\pi$ phase coverage and high transmission (unity in the ideal case) in a single layer structure. In contrast to the tall dielectric pillar based metasurfaces~\cite{vmp2018,amin2019}, the dielectric metasurfaces based on the interference of Mie resonances (Huygens' metasurface)~\cite{as2021,md2015} are low-profile (much flatter) in nature.

Despite their tremendous potential, DHMS are more popular across the optical regimes, presenting a wide range of applications, including anomalous refraction~\cite{ajo2018,cl2019,liu2020,yfy2015,ao2017}, focusing~\cite{ao2017,yt2020}, holography~\cite{chong2016,wz2016}, bessel beam generation~\cite{zl2019}, to mention a few. Such meta-devices are also of the utmost importance for microwave and millimeter-wave applications, such as beamforming, polarization control, gain enhancement, among many others. Nonetheless, the capabilities of dielectric metasurfaces, including DHMS, are not much explored in the microwave and mm-wave regimes, and only a handful of designs have been reported~\cite{vmp2018,amin2019,ka2017,am2020,mk2020,mke2020,emara2021}. The implementation of wave-deflection meta-devices in~\cite{ajo2018,cl2019,vmp2018,liu2020,yfy2015,mk2020,mke2020,emara2021} employs the phase gradient approach based on the generalized Snell's laws~\cite{nyu2011}. The meta-devices realized based on the phase-gradient technique fundamentally suffer from low efficiencies, especially at the extreme steering angle, due to the impedance mismatch between the incident and scattered waves~\cite{nme2016,vsa2016}. In addition, the metasurface requires high-resolution discretization, comprising a large number (10-20) of subwavelength sized elements per wavelength to mimic the continuous phase profile. The metasurfaces with a dense distribution of subwavelength sized meta-atoms possess strong mutual coupling between the elements and are subject to implementation difficulties, particularly at higher frequencies. To address some of the shortcomings of phase-gradient metasurfaces, a design approach based on surface impedance -- satisfying generalized sheet transition conditions (GSTCs), has been introduced to realize efficient meta-devices~\cite{mc2018,chen2020}. Nevertheless, similar to phase-gradient metasurfaces, implementing rapidly varying surface impedance profile in GSTCs-oriented designs still necessitates dense discretization.

Recently, the idea of \emph{coarse discretization}~\cite{amhw2018,awong2018} has been put forward in pursuit of extreme EM wave manipulation with high efficiency and simplicity. A coarsely discretized metasurface design adopts a different approach that takes advantage of classical grating physics as opposed to gradient metasurfaces based on the generalized Snell's law. The gradient metasurface introduces transverse spatial modulation in terms of phase or surface impedance, locally imparting an additional momentum to transform the impinging wavefront. In the case of the discretized metasurface, the periodicity selects a discrete set of Floquet channels, known as \emph{Floquet-Bloch} (FB) modes -- a series of propagating and evanescent plane waves excited by periodic structure. The metasurface periodicity is set to redirect an impinging wave towards one of the propagating higher-order FB mode rather than imposing an additional transverse momentum. Other propagating FB modes may exist (at the very least the fundamental mode \emph{i.e} $0^{\text{th}}$ order mode which represents the specular reflection and direct transmission always exists) and these modes are suppressed by carefully engineering the scatterers. Coarse discretization significantly reduces the number of polarizable particles in a period with a relatively large but still sub-wavelength unit-cell size, which minimizes mutual coupling between the elements and simplifies fabrication~\cite{as2021,amhw2018,awong2018,sharma2020,asharma2020,aswong2021,qic2022,qi2022,sw2022}. Coarse discretization also reduces the number of propagating modes which greatly helps control and suppress spurious scattering.

At this stage, it is important to distinguish between a coarsely discretized metasurface and a metagrating~\cite{radi2017,yradi2018,rab2018,tan2019,vp2020,xu2021,vkk2021}. The key points are summarized next, but the reader can refer to~\cite{qic2022} for additional information. On a fundamental level, both metagrating and the coarsely discretized metasurface use the grating physics approach as described above. At the implementation level, metagrating differs from our coarse discretization approach in a way that the former views the period as a single unit, consisting of a complex single~\cite{radi2017,rab2018} or multiple~\cite{yradi2018} polarizable particles to engineer the surface for wavefront manipulation, whereas the latter offers a more specific and streamlined approach for controlling the FB modes by correlating the number of modes to the number of elements required per period for complex wave transformation.

Wong and Eleftheriades in~\cite{amhw2018} have suggested that by discretizing the period with \emph{N}-meta-atoms is adequate to control the \emph{N}-Floquet modes which propagate into the far-field. The judicious tuning of the individual meta-atoms, suppresses the spurious FB modes, enhancing the power in the desired propagation mode, which can scatter in a specific direction, dictated by the metasurface periodicity and the angle of incidence. The discretization within the period to accommodate multiple scatterers enables sophisticated design strategies to modulate the required FB modes to realize complex wavefront transformation.

Recognizing the potential of coarse discretization, we present in this paper a simple, cost-effective, efficient and robust dielectric Huygens' metasurface at mm-wave frequency. The proposed metasurface features only two rectangular dielectric meta-atoms per period, and hence it is termed as \emph{bipartite dielectric Huygens' metasurface} (Bi-DHMS). As an example, we show near-reflectionless anomalous refraction at 28 GHz (useful for 5G mm-wave applications), where an incoming linearly polarized wave at $\theta_i=15\degree$ redirects towards $\theta_t=-44.5\degree$ with the refraction efficiency of 87\%. Focusing on engineering the \emph{FB}-mode rather than employing the conventional phase-gradient approach, we arrive at a design that is a combination of high angular deflection and high power efficiency unmatched by most dielectric metasurfaces~\cite{ajo2018,cl2019,liu2020,yfy2015,ao2017,mk2020,mke2020,emara2021}. Additionally, a systematic design approach is presented in this paper which allows the design to be easily scaled to different parts of the electromagnetic spectrum.
\section{Dielectric Huygens' Metasurface}
\begin{figure}[t]
	\centering
	\includegraphics[scale=0.6]{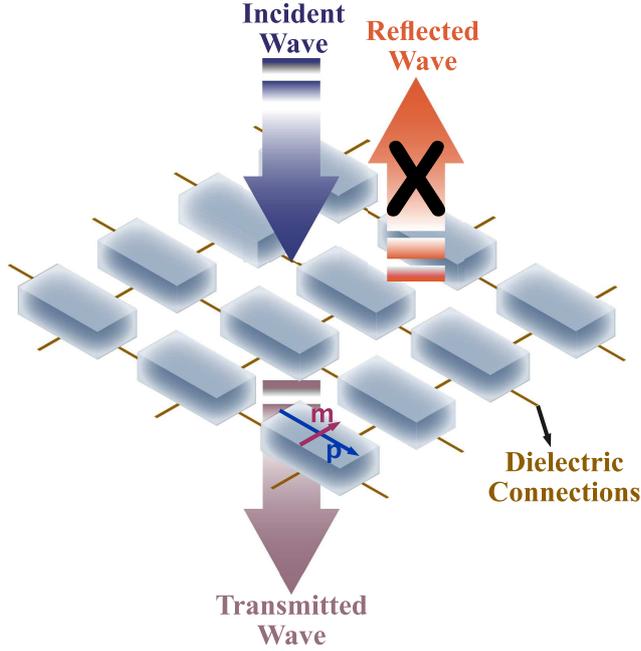}
	\caption{\label{fig1} A generic illustration of dielectric Huygens' metasurface formed by 2D array of rectangular dielectric particles, each representing spectrally overlapped crossed electric dipole ($\mathbf{p}$) and magnetic dipole ($\mathbf{m}$).}
\end{figure}

\begin{figure}[!t]
	\centering
	\includegraphics[scale=0.9]{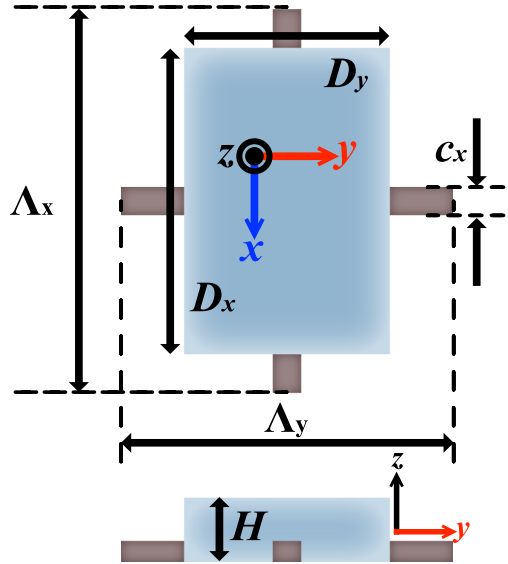}
	\caption{The top and side view of the proposed dielectric meta-atom. The unit-cell dimensions are: $\Lambda_x=0.52\lambda_0$, $\Lambda_y=0.75\lambda_0$, $c_x=\lambda_0/21$, $L_x=5.5$, $L_y=4$, and $H=1.45$ (Unit: mm). The free space wavelength $\lambda_0$ is calculated at 28 GHz.}\label{f2}
\end{figure}

Figure \ref{fig1} shows a generic illustration of the dielectric Huygens' metasurface (DHMS), comprising a 2-D array of rectangular dielectric resonators (DRs) as meta-atoms. Each meta-atom is defined by a pair of spectrally overlapped crossed electric dipoles (ED, $\mathbf{p}$) and magnetic dipoles (MD, $\mathbf{m}$), which act as a Huygens' source, providing unidirectional scattering and canceling out any reflection from the metasurface.
The rectangular dielectric meta-atom ($\epsilon_r=12$), shown in Figure \ref{f2}, serves as the fundamental entity of the proposed metasurface, which, in comparison to the cylindrical particle, offers an extra degree of freedom to engineer the spectral overlap of the ED and MD modes. Note that the meta-atom consists of four deep subwavelength-sized dielectric connections (having same relative permittivity as that of the rectangular dielectric blocks) to hold the dielectric resonator in a periodic array. These connections may slightly perturb the field; however, this can be compensated through proper optimization.

\begin{figure*}[!t]
	\centering
	\subfloat[]{
		\quad
		\includegraphics[scale=0.45]{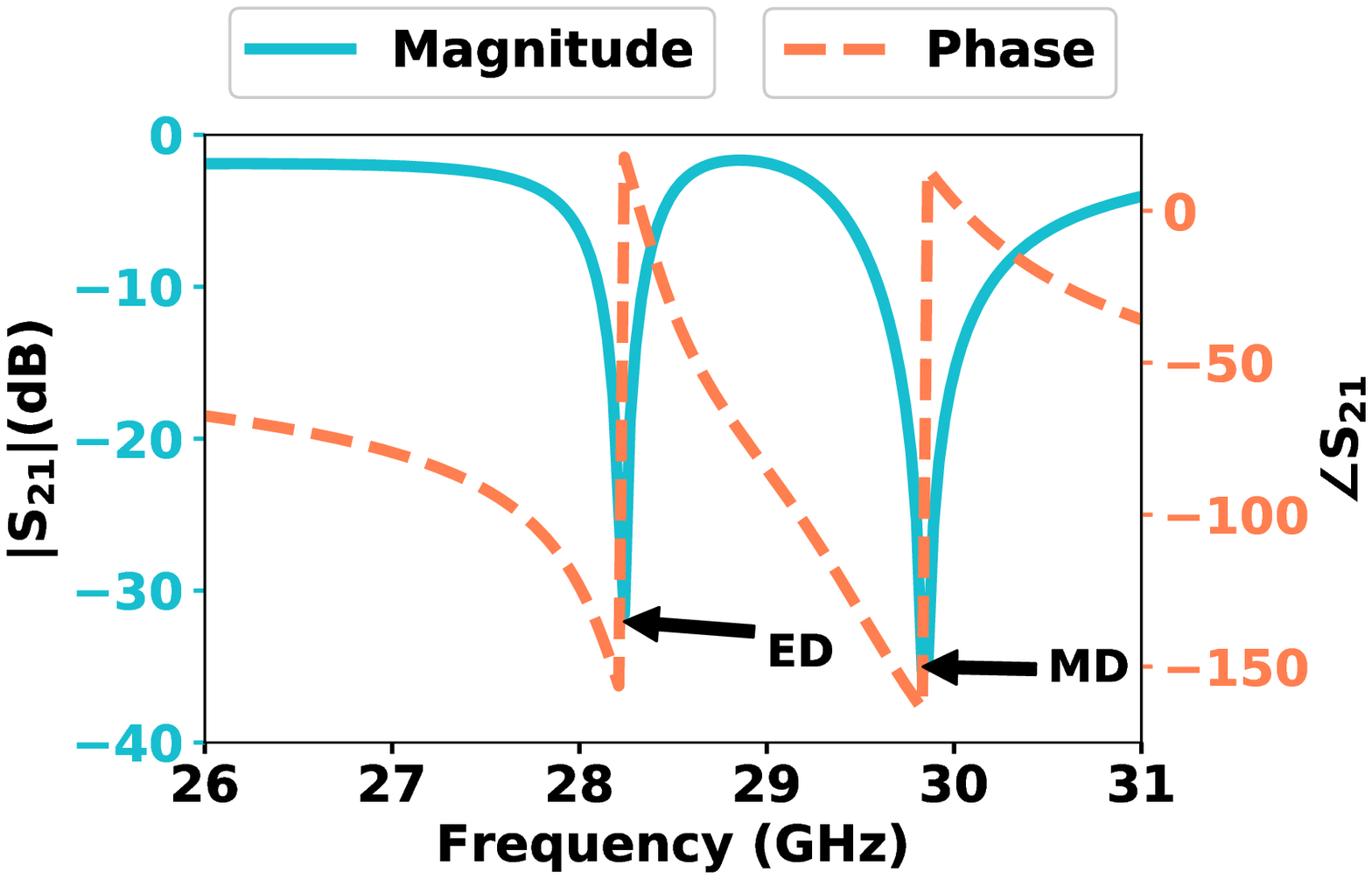}}
	\subfloat[]{\raisebox{4.5ex}{
			\includegraphics[scale=0.35]{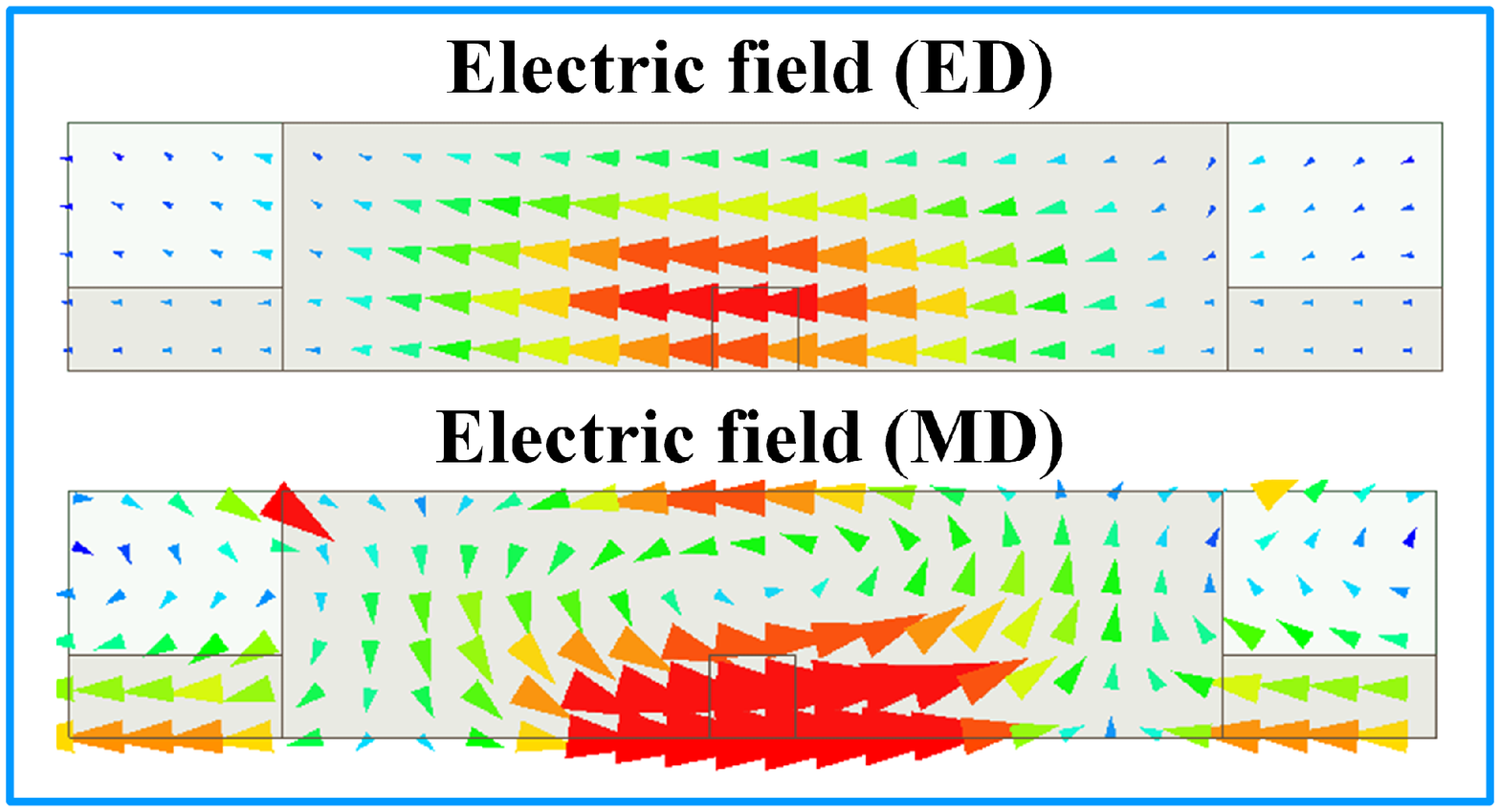}}}
	\caption{\label{fig3} Simulation of rectangular dielectric meta-atom placed in an infinite periodic array along the \emph{x}- and \emph{y}-directions. (a) Transmission spectra (magnitude and phase), where ED represents electric dipole (28.2 GHz) and MD represents magnetic dipole (29.8 GHz). (b) Electric field distribution in the \emph{xz}-plane. The upper panel shows electric field lines at 28.2 GHz that resembles an ED aligned in the \emph{x} direction and the lower panel depicts electric field lines at 29.8 GHz which resembles an MD oriented along the \emph{y} direction.}
\end{figure*}

We start with the simulation of the dielectric meta-atom (refer to Figure \ref{f2}), placed in an infinite array along the \emph{x}- and \emph{y}-directions, to analyze the electric and magnetic dipolar modes.
The simulation is carried out using Ansys HFSS by imposing the periodic boundary conditions along \emph{x}- and \emph{y}-directions and using Floquet ports to excite the structure by \emph{x}-polarized (electric field points along \emph{x}-direction) plane wave propagating in the negative \emph{z}-direction. The relative permittivity of the dielectric material is 12, and the dielectric loss is not considered at this stage. Figure \ref{fig3}(a) depicts the transmission spectra (magnitude and phase) of a metasurface built from the periodic repetition of an example dielectric meta-atom, showing two transmission magnitude minima at 28.2 GHz and 29.8 GHz. Correspondingly, the phase response shows a maximum phase change of $180\degree$ for both resonances. 
Figure \ref{fig3}(b) depicts the vector electric field distribution in the \emph{xz}-plane at the resonant frequencies mentioned above. The examination of electric field lines at 28.2 GHz reveals that the ED mode is indeed excited, orienting along \emph{x}-direction. On the other hand, at 29.8 GHz, the electric field lines are anti-parallel at opposite sides, indicating the excitation of the MD mode oriented along \emph{y}-direction. The ED and MD resonances, in this case, are well separated and distinguishable; however, by appropriately adjusting the geometrical parameters of the rectangular DR, these modes can interfere constructively, leading to high transmission. Next, we demonstrate how changing the geometrical parameters of the unit-cell ($\Lambda_x$, $\Lambda_y$, $D_x$, $D_y$, and $H$) can affect the ED and MD mode resonance.

We study the effect of varying unit-cell parameters on the ED and MD resonances of the DR arranged in an infinite periodic array. When illuminated by an \emph{x}-polarized incident plane wave, two orthogonal dipole modes, \emph{viz.} the electric dipole ($\mathbf{p_x}$) and the magnetic dipole ($\mathbf{m_y}$), are excited, and these modes are aligned in the \emph{x}- and \emph{y}-directions, respectively.

\begin{figure}[t]
	\centering
	\includegraphics[scale=0.7]{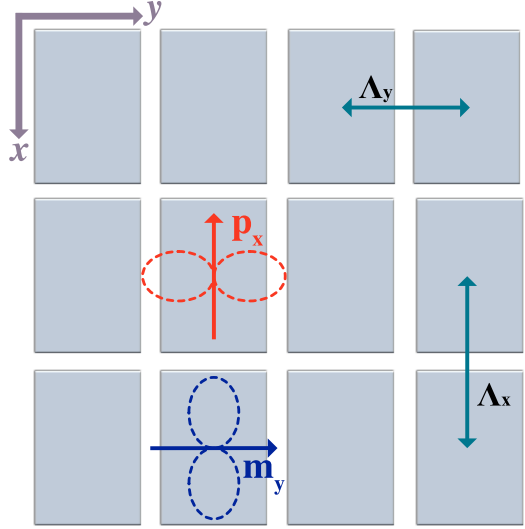}
	\caption{\label{fig2} Rectangular dielectric resonators arranged in an infinite periodic array with periodicity $\Lambda_x$ and $\Lambda_y$ in the \emph{x}- and \emph{y}-directions.}\label{s1}
\end{figure}

We first show how the magnetic and electric dipole resonances can be engineered separately by varying the periodicity in only one of the two orthogonal directions in the plane \emph{i.e.} $\Lambda_x$ or $\Lambda_y$ (refer to Figure \ref{s1}).
\begin{figure*}[h]
	\centering
	\subfloat[]{
		\includegraphics[scale=0.6]{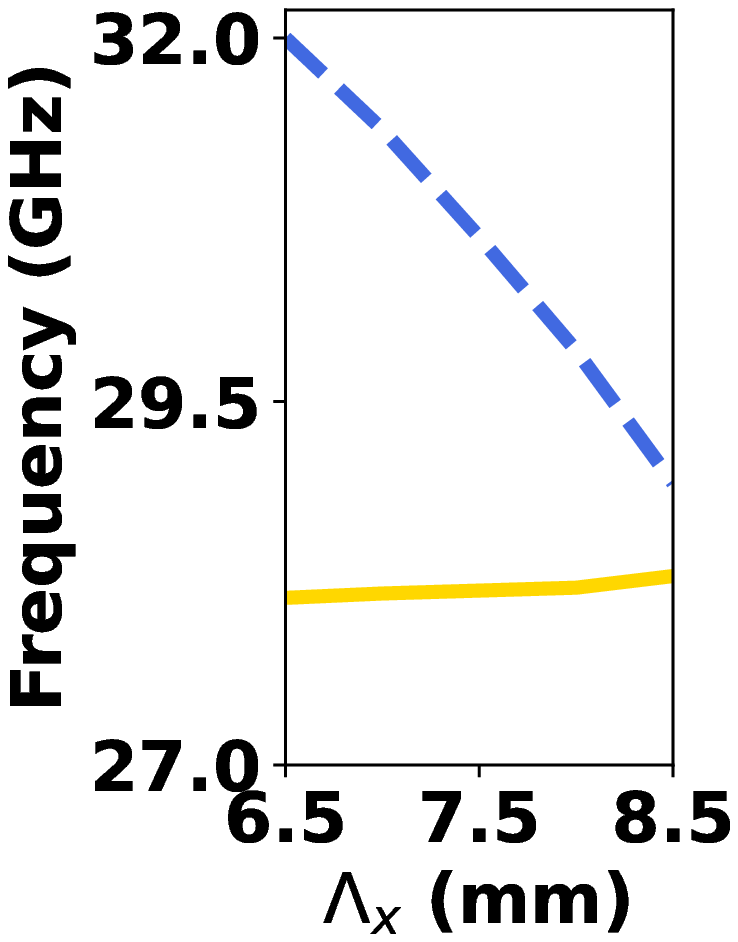}}
	\subfloat[]{
		\includegraphics[scale=0.6]{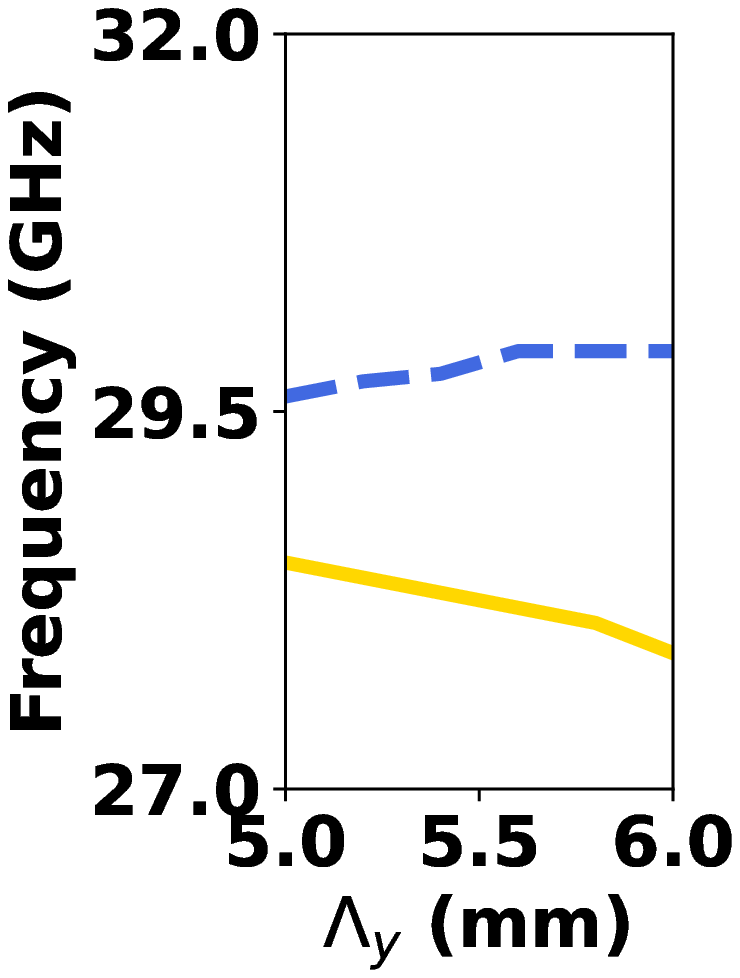}}
	\caption{Plots showing the change in electric (solid) and magnetic (dashed) dipole resonance positions with the variation in (a) $\Lambda_x$ and (b) $\Lambda_y$.}\label{s2}
\end{figure*}
\begin{figure*}[t]
	\centering
	\subfloat[]{
		\includegraphics[scale=0.6]{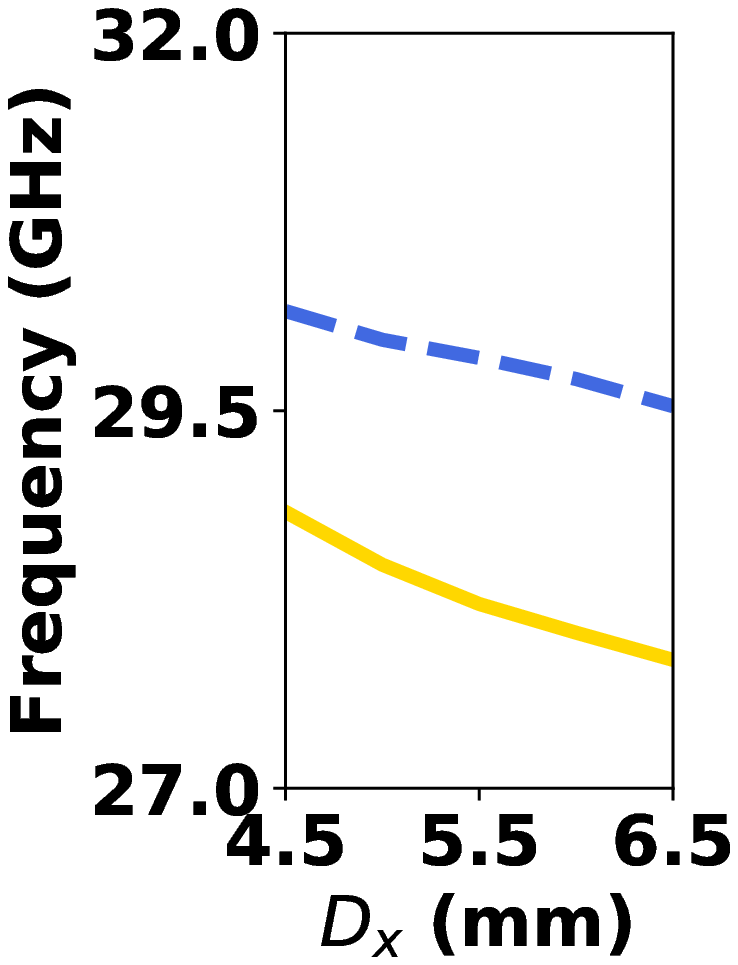}}
	\subfloat[]{
		\includegraphics[scale=0.6]{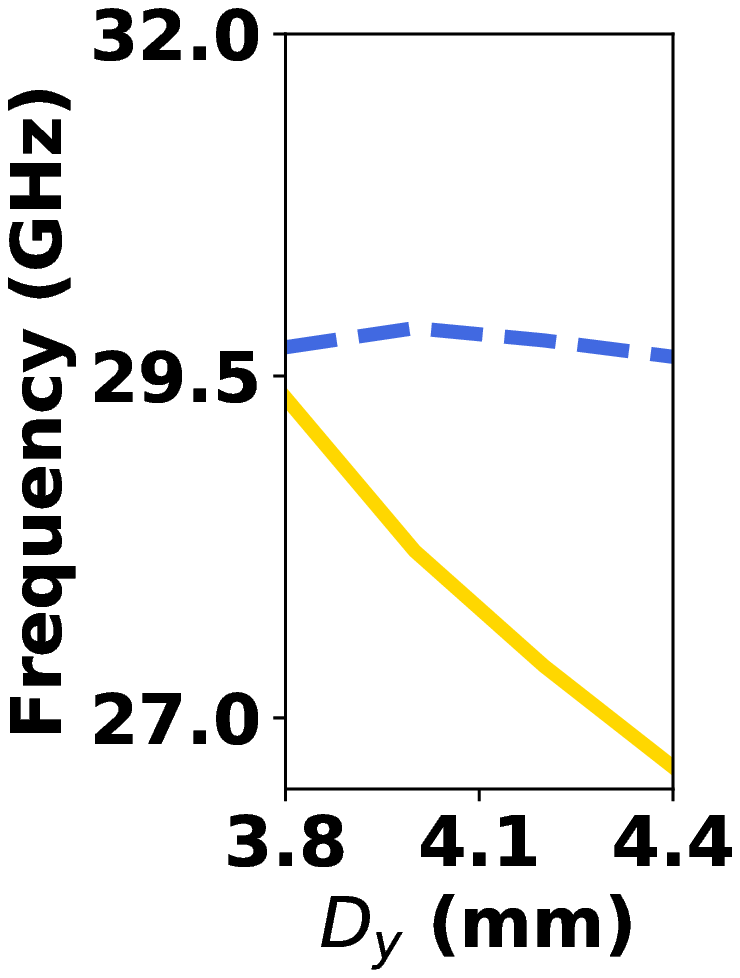}}
	\subfloat[]{
		\includegraphics[scale=0.6]{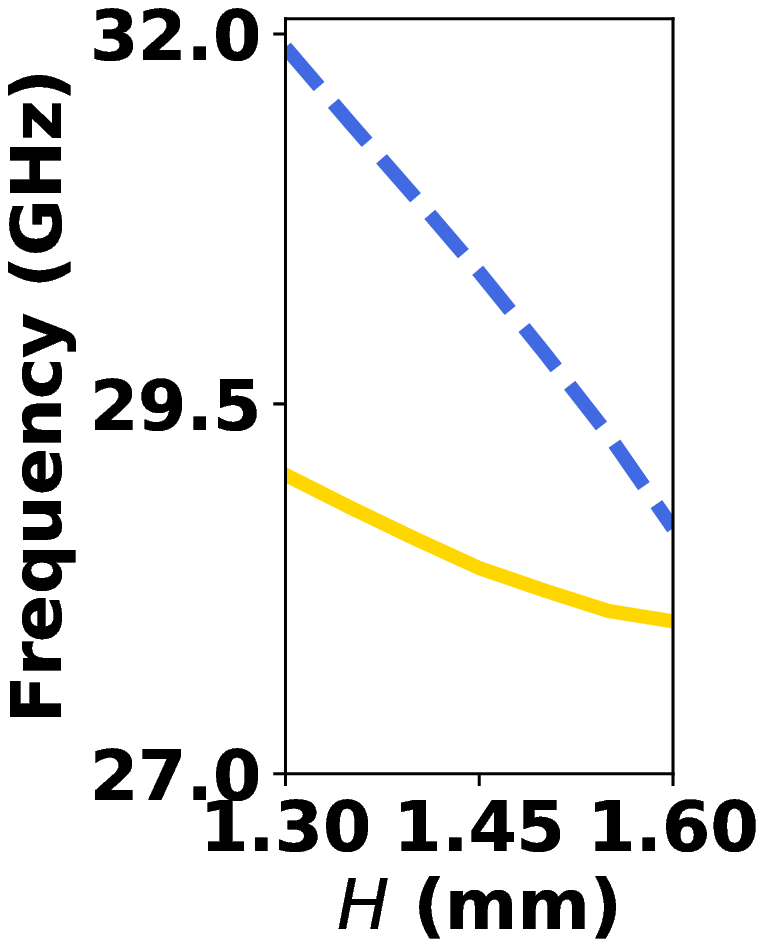}}
	\caption{Plots showing the change in electric (solid) and magnetic (dashed) dipole resonance positions with the variation in (a) $D_x$ (b) $D_y$ and $H$.}\label{s3}
\end{figure*}

As depicted in Figure \ref{s1}, the direction of maximum radiation for an MD is along the \emph{x}-axis (refer to the blue dashed line in the Figure \ref{s1}) and it has a far-field null along the \emph{y}-axis. The former means that the MD experiences strong mutual coupling in the \emph{x}-direction and is susceptible to a variation in $\Lambda_x$. The latter means that the EM field decays rapidly in the \emph{y}-direction away from the DR. As a result, the $\mathbf{m_y}$ mode experiences relatively weak mutual coupling in the \emph{y}-direction, and is hence less affected by a variation in $\Lambda_y$. Periodic simulations, depicted in Figures~\ref{s2}(a)-(b), verify that the MD resonance is affected strongly by a variation in $\Lambda_x$ (while $\Lambda_y$ is kept constant) and weakly by a variation in $\Lambda_y$ (while $\Lambda_x$ is kept constant). The reverse is true for $\mathbf{p_x}$. Our finding is in agreement with conclusions from previous works on cylindrical DRs~\cite{veb2018,jia2018}; further, in this work we advocate the rectangular DR as it has more degrees of freedom than the cylindrical one. With the more degrees of freedom and the formation of periodic array with carefully engineered rectangular periodicity, the design offers a more promising potential to accomplish a wide range of functionalities at mm-wave frequencies and above.

Apart from varying the periodicity of the array, the electric and magnetic dipole resonance can be engineered by changing the dimensions of the dielectric resonator. Next, we show how the geometrical parameters of rectangular DR affect the ED and MD mode resonances. Because all of the dimensions determine the resonant frequency of the DR ($D_x$, $D_y$, and $H$)~\cite{pan2011}, changing these dimensions (while keeping $\Lambda_x$ and $\Lambda_y$ constant) affects both resonant frequencies, as depicted in Figures~\ref{s3}(a)-(c). Based on the above observations, one can realize the spectral overlap by judiciously tuning the geometrical parameters of the dielectric meta-atom.

\begin{figure*}[!t]
	\centering
	\subfloat[]{
		\includegraphics[scale=0.5]{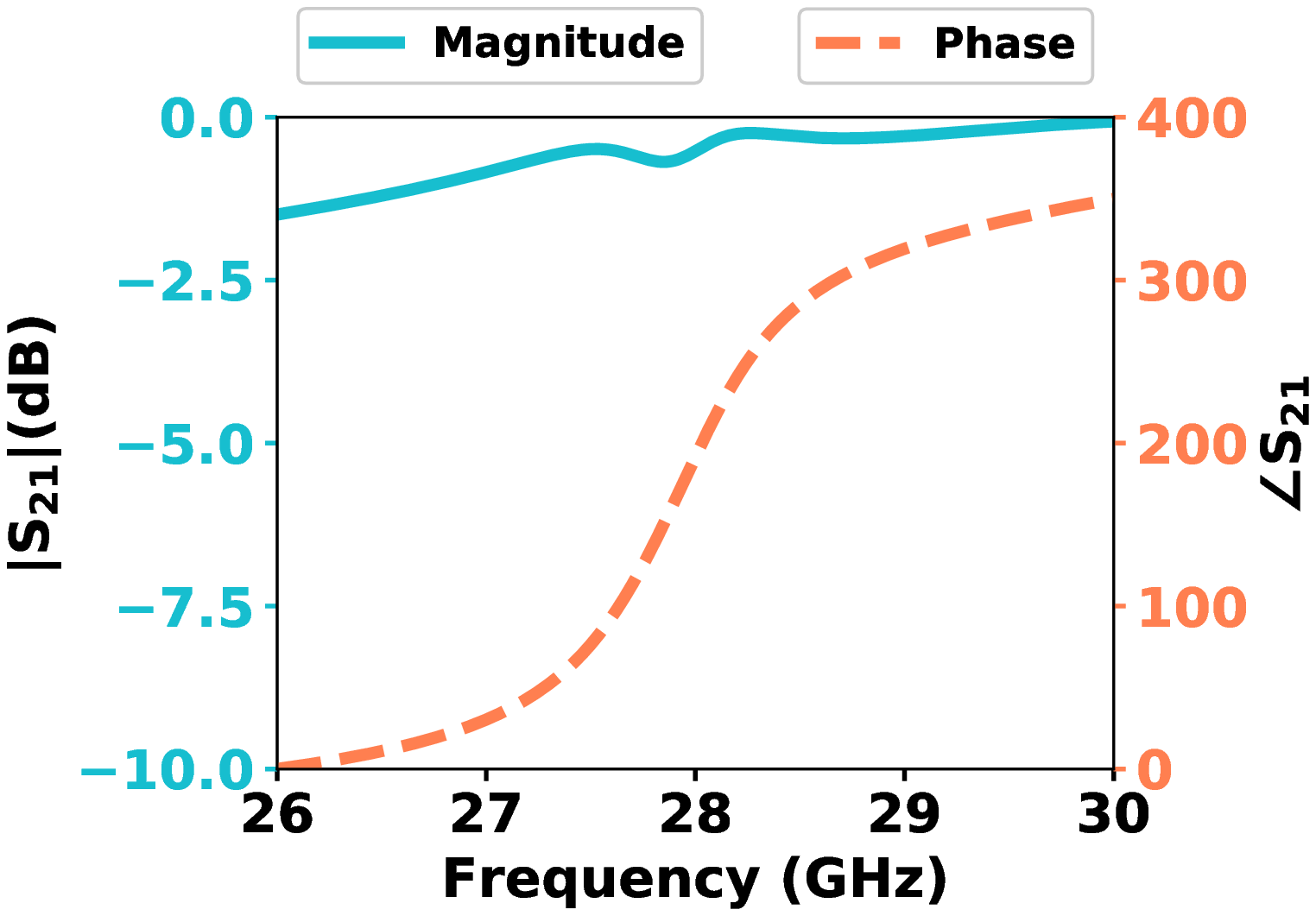}}
	\subfloat[]{
		\includegraphics[scale=0.5]{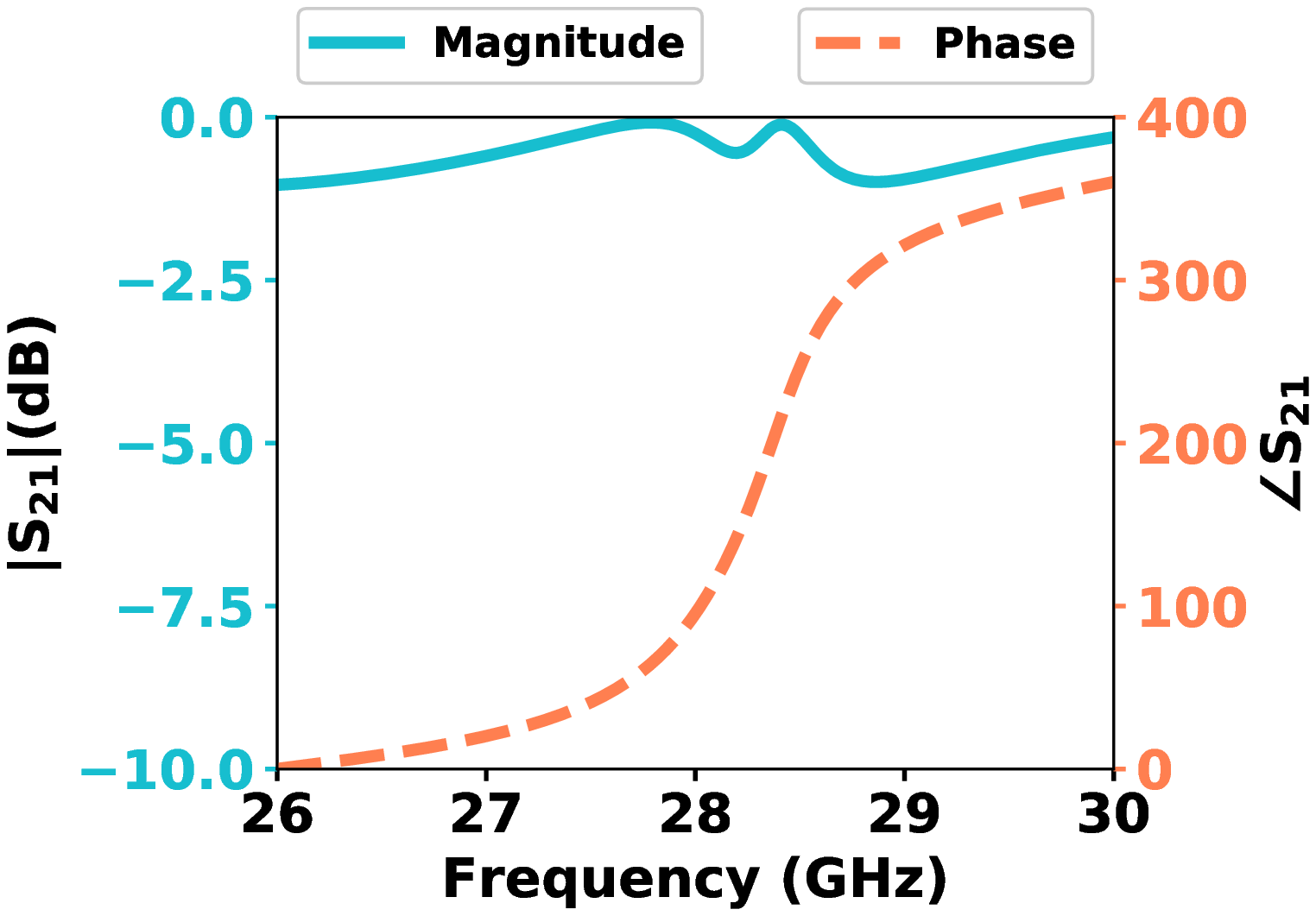}}
	\caption{\label{fig5} (a) Transmission spectra of dielectric Huygens' meta-atom for normal incidence, showing high transmission coefficient ($|\text{S}_{21}|>-1$ dB) and $2\pi$ phase coverage across 26-30 GHz. (b) Transmission spectra of dielectric Huygens' meta-atom for $15^{\circ}$ oblique incidence, achieving high transmission coefficient ($|\text{S}_{21}|>-1$ dB) with full $2\pi$ phase coverage.}
\end{figure*}

We now optimize the meta-atom geometry to realize the spectral overlapping of the above-mentioned electric and magnetic dipole modes. Figure \ref{fig5}(a) shows the transmission magnitude and phase profiles of the spectrally overlapped dipolar modes. 
In contrast to separate resonances, we here observe a high transmission ($|\text{S}_{21}|>-1$ dB) across 26-30 GHz. More importantly, the transmission phase ($\angle\text{S}_{21}$) spans the entire $2\pi$ phase range, which is double the phase change that can be obtained by single electric or magnetic resonance. In Ref.~\cite{da2017}, it has been demonstrated for cylindrical DRs placed in an infinite periodic array that the transmission response for both separate and overlapped resonances show dependence on the incident angle. For instance, in the case of spectrally overlapped resonances, while the metasurface is almost transparent across the entire spectral range under normal incidence, a gradual decrease in the transmission has been observed as the incidence angle increases. We have also observed a similar phenomenon for rectangular DR under an off-normal incidence. Therefore, to regain full transparency across the spectral range for the incident angle of interest, which in our case is $15^{\circ}$, we fine-tune the dimensions of the rectangular DR. The transmission spectra of the re-optimized meta-atom for $\theta_i=15^{\circ}$ is shown in Figure \ref{fig5}(b). In this case, the metasurface is almost transparent over the entire spectral range of interest (26-30 GHz) and achieves full $2\pi$ phase coverage.
\section{Metasurface Design Methodology}
\subsection{Coarse Discretization}
\begin{figure}[t]
	\centering
	\includegraphics[scale=0.5]{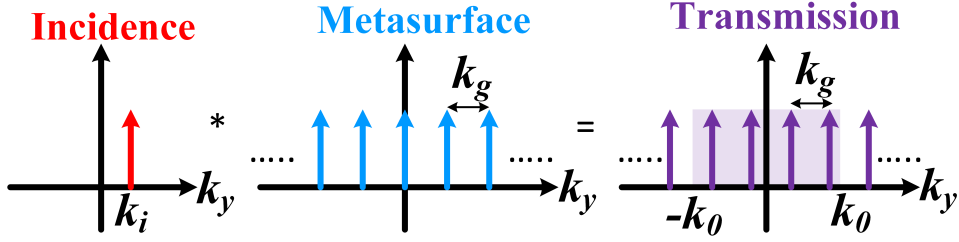}
	\caption{\label{fig6}A diagram showing the $k$-space operation of a periodic metasurface which varies along $y$-direction. The arrows represent the existence of Floquet-Bloch modes (not the amplitude and phase). The purple box in the rightmost figure depicts the propagation regime, with $k_y \in [-k_0,k_0]$.}
\end{figure}

The coarsely discretized metasurface's design principle is to coalesce the classical grating physics with the capabilities of the metasurface, allowing extreme manipulation of EM waves with simplicity and efficiency. Consider a periodic metasurface in free-space with spatial frequency $k_g$, and periodicity $\Lambda_g=\frac{2\pi}{k_g}$. When a plane wave illuminates such a structure, a series of propagating and evanescent plane waves -- referred to as \emph{Floquet-Bloch} (FB) modes, are excited. The transverse spatial frequency of these modes is expressed as
\begin{equation}\label{eq1}
	k_{y(n)}=k_{i}+nk_g=k_i+n\frac{2\pi}{\Lambda_g},
\end{equation}
where $k_i$ represent transverse $y$-directed wavenumber of the incident wave. Figure \ref{fig6} shows the spectral domain (\emph{k}-space) operation of a periodic metasurface, which can be expressed mathematically as
\begin{equation}\label{eq2}
	\Omega_o(k_y)=\sum_p A_n \delta(k_y-k_{y(n)}),
\end{equation}
where $\Omega_o(k_y)$ is the the \emph{k}-space spectrum of the transmitted EM wave, $A_n$ represents the complex amplitude, and integer $n=0,\;\pm1,\;\pm2,\;\cdots$ represent the $n^{th}$ FB mode.
The arrows in Figure \ref{fig6} show the presence of spectral components; however, they do not represent the amplitude and phase of these spectral components. The local period of the metasurface selects a discrete set of FB modes that lie in the propagation regime, with $k_y \in [-k_0,k_0]$ (depicted by the purple box in Figure \ref{fig6}), which can scatter in the specific allowed direction, determined by the incidence angle and periodicity of the metasurface. The modes that lie outside the aforementioned propagation regime are evanescent and will remain in the near-field of the metasurface. Wong and Eleftheriades have shown that \emph{N}-fold discretization within the metasurface period is adequate to manipulate the power carried by the \emph{N}-FB modes that can scatter into the far-field~\cite{amhw2018}. Leveraging this principle, the authors have successfully demonstrated simple metasurface designs for perfect anomalous reflection~\cite{amhw2018} and retroreflection~\cite{awong2018}.
\subsection{Propagating Modes}
\begin{figure}[t]
	\centering
	\includegraphics[scale=1]{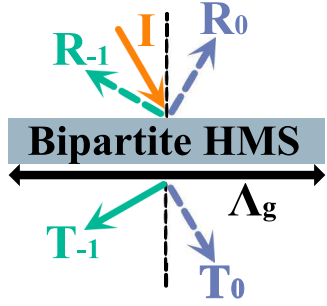}
	\caption{\label{fig7} A diagram illustrating all the FB modes (transmission and reflection) for the metasurface. Orange solid arrow represent the incident wave (I) and the green solid arrow indicate the desired FB mode ($n=-1$) on the transmission side. The arrows with dashed line represent the undesired modes, where $\text{T}_0$ represent direct transmission and $\text{R}_0$ and $\text{R}_{-1}$ denotes specular and first order reflection.}
\end{figure}

The FB modes that fall under the propagation regime scatters into specific allowed directions $\theta_{t(p)}$, given as
\begin{equation}\label{eq4}
	\sin\theta_{t(n)}=\sin\theta_i+\frac{n \lambda_0}{\Lambda_g},
\end{equation}
where $\theta_i$ is the angle of incidence, $n$ represents the FB mode number, and $\Lambda_g$ is the periodicity of the metasurface. Our goal is to design a coarsely discretized metasurface that effectively couples the incident power to the desired $n=-1$ mode departing at an angle of $\theta_{t(-1)}$. To ensure that only the $n=0$ and $n=-1$ (anomalous) modes reside in the propagation regime~\cite{radi2017,rab2018}, the metasurface periodicity should satisfy
\begin{eqnarray}\label{eq5}
	\frac{\lambda_0}{1+\sin\theta_i} < \Lambda_d < \left\{\begin{array}{r@{\quad}cr} 
		\frac{\lambda_0}{1-\sin\theta_i} & 0 < \theta_i< \sin^{-1}(1/3) \\  
		\frac{2\lambda_0}{1+\sin\theta_i} & \sin^{-1}(1/3)<\theta_i<\frac{\pi}{2}  
	\end{array}\right.
\end{eqnarray}
Equation (\ref{eq5}) is mapped onto the angular domain as
\begin{equation}\label{eq6}
	\small
	\frac{-\pi}{2} < \theta_t < \left\{  \begin{array}{r@{\quad}cr} 
		\sin^{-1}(2\sin\theta_i-1)  & 0 <\theta_i<\sin^{-1}(1/3) \\
		\sin^{-1}\left(\frac{1}{2}\sin\theta_i-\frac{1}{2}\right) & \sin^{-1}(1/3)<\theta_i<\frac{\pi}{2}
	\end{array}\right.
\end{equation}
It should be noted that (\ref{eq5}) and (\ref{eq6}) are valid for $\theta_i>0$. In the case of the transmissive metasurface, the meta-atoms are engineered to channel most of the incident power into $n=-1$ transmission mode while suppressing direct transmission and both specular and first-order reflections, as illustrated in Figure \ref{fig7}.
\subsection{Bipartite Dielectric Huygens' Metasurface: Design and Simulation}
As a proof of concept, we design a transmissive dielectric Huygens' metasurface at 28 GHz that reroutes a wave impinging from $\theta_i=15\degree$ to the first order FB mode ($n=-1$) departing at an angle $\theta_t=-44.5\degree$. For this case, (\ref{eq4}) dictates the period the metasurface along $y$ direction as $\Lambda_g=\frac{\lambda_0}{|\sin\theta_i-\sin\theta_{t(-1)}|}=1.04\lambda_0$. In our case, a total of four FB modes \emph{viz.} $\text{T}_{0}$, $\text{T}_{-1}$, $\text{R}_{0}$, and $\text{R}_{-1}$, where $\text{T}$ denotes transmission and $\text{R}$ denotes reflection, exists in the propagation region. Fortunately, the nature of Huygens' metasurface is such that it produces extremely weak reflections and can be considered as a near-reflectionless artificial surface~\cite{mc2018,chen2020}. As a result, we are left with only two modes in the transmission regime, and hence a binary discretization -- two meta-atoms per period, will suffice to manipulate the power carried by the aforementioned FB modes.
\begin{figure}[t]
	\centering
	\includegraphics[scale=0.5]{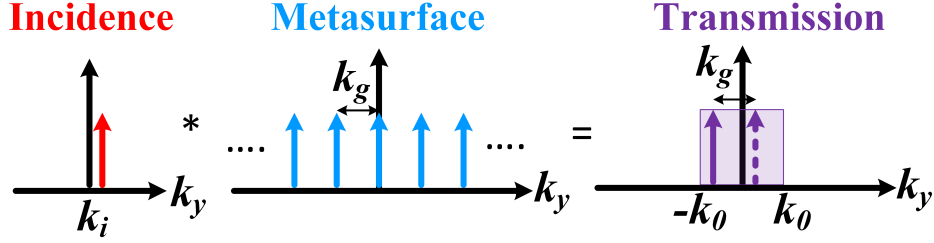}
	\caption{\label{fig8} The \emph{k}-space operation of a periodic metasurface with binary discretization level -- two elements per period, under oblique incidence. The arrows inside the purple box depict the transmitted \emph{Floquet-Bloch} modes, where, the solid arrow represents the mode with non-zero amplitude and dashed arrow indicates the mode having zero amplitude. The evanescent modes are not depicted.}
\end{figure}
Figure \ref{fig8} shows the transformation of a plane wave's \emph{y} directed wave vector as it refracts off the metasurface.

Before designing the metasurface for anomalous refraction, we create a \emph{library of meta-atoms} having different transmission phases and similar transmission magnitude by varying the geometrical parameters ($D_x$, $D_y$, and $H$) of the rectangular DR placed in an infinite periodic array along the $x$- and $y$-directions. 
\begin{figure}[!t]
	\centering
	\subfloat[]{
		\includegraphics[scale=0.8]{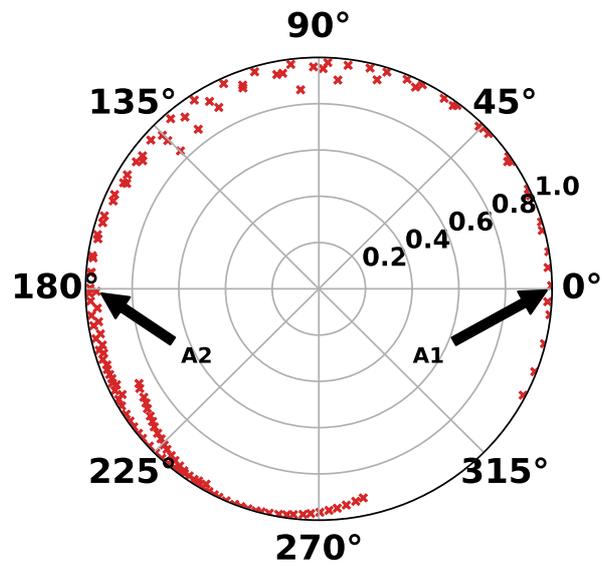}}\\
	\subfloat[]{
		\includegraphics[scale=0.8]{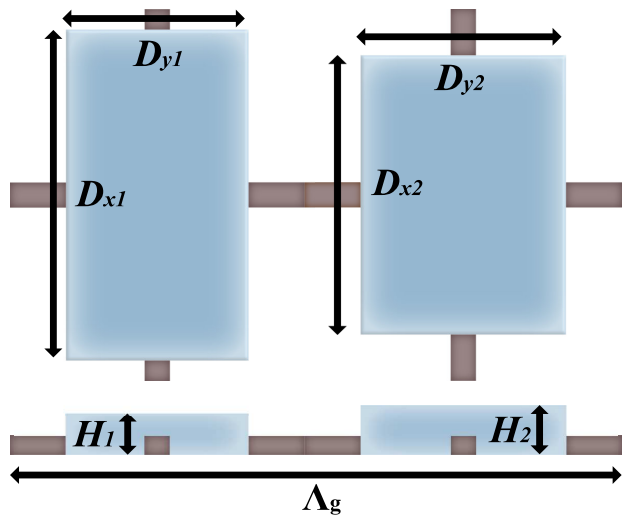}}
	\caption{\label{fig9} Design and simulation of bipartite dielectric Huygens' metasurface. (a) A plot showing the transmission magnitude and phase for different combinations of $D_x$, $D_y$, and $H$. (b) A schematic of one period (supercell) of the proposed metasurface.}
\end{figure}
The one period of the metasurface is then built by selecting two elements from the library (refer to arrows A1 and A2 in Figure \ref{fig9}(a)) with near-unity transmission and a phase difference of $180\degree$. We call the proposed metasurface a \emph{bipartite dielectric Huygens' metasurface} -- a metasurface comprising two elements per period. The mutual interaction dynamics between these adjacent elements impact the transmission properties, degrading the metasurface performance. In light of this, we fine-tune the resonator dimensions in the new simulation environment to restore suppression of unwanted FB modes.Figures \ref{fig10} (a) and \ref{fig10} (b) show the transmittance and reflectance for the propagating FB modes, respectively.
\begin{figure}[t]
	\centering
	\includegraphics[scale=0.8]{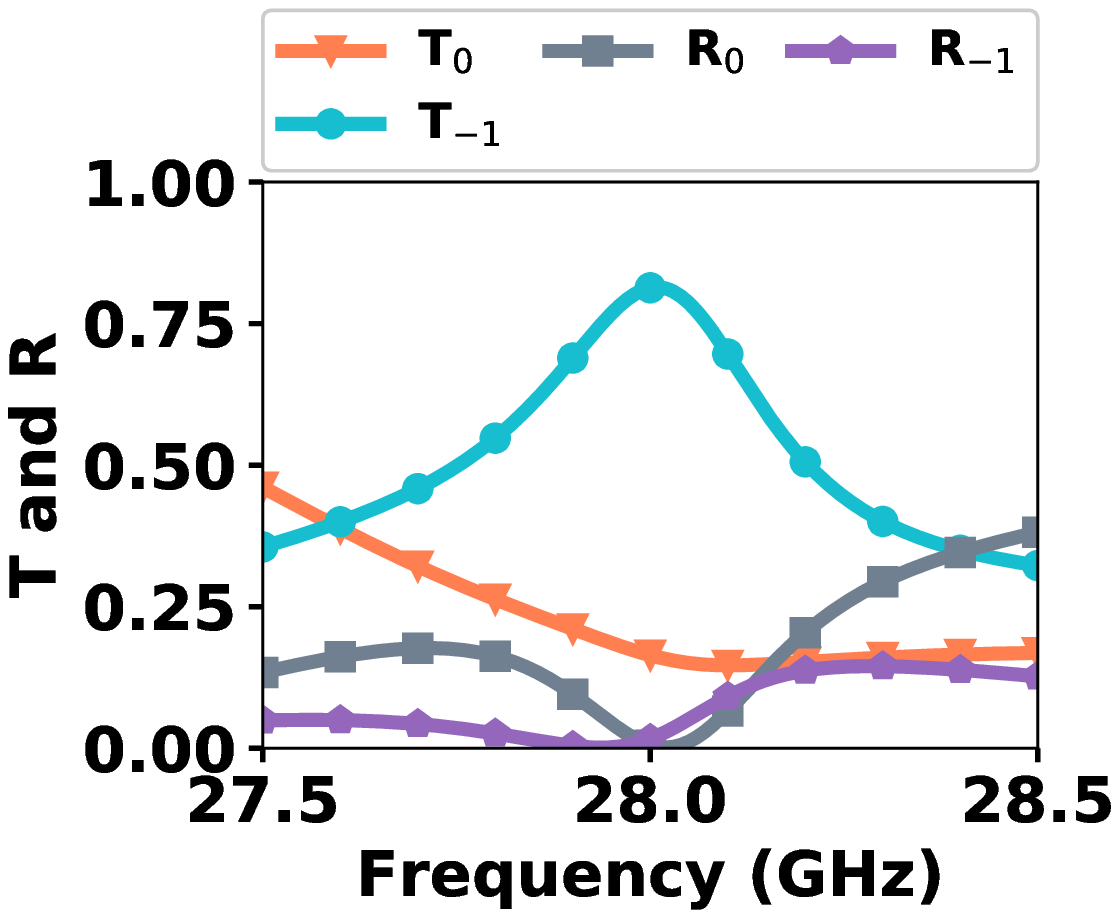}
	\caption{\label{fig10} Transmittance (T) and reflectance (R) of the -1 and $0^{\text{th}}$ FB modes propagating towards $-44.5\degree$ and $0\degree$, respectively.}
\end{figure}
\begin{figure}[h]
	\centering
	\includegraphics[scale=0.6]{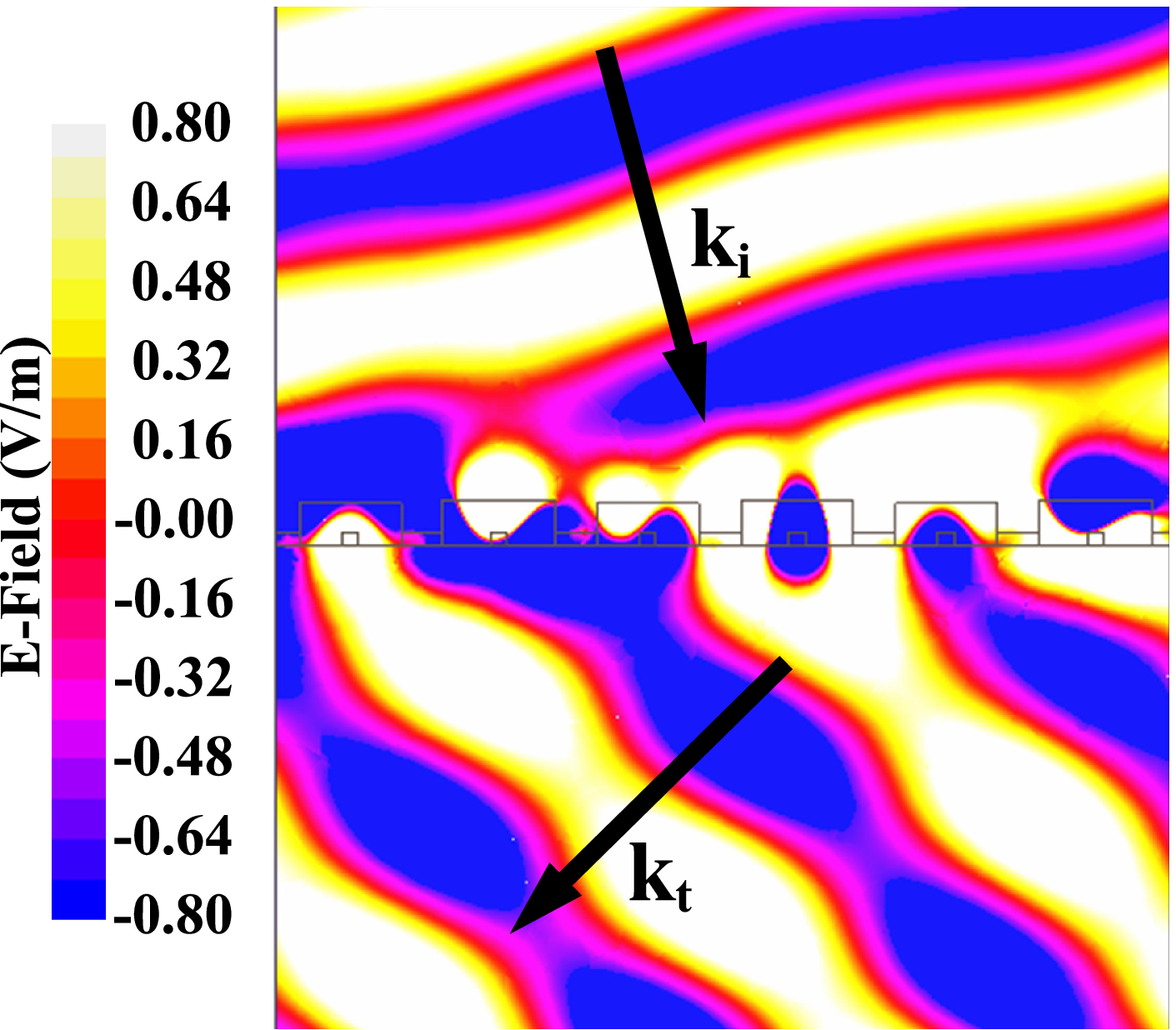}
	\caption{\label{fig11}Electric field distribution in \emph{yz}-plane at 28 GHz, showing anomalous refraction. The upper region of the metasurface shows the incident wave impinging at an angle $\theta_i=15\degree$ with minimal reflection, whereas the lower region clearly depicts the refracted wave front.}
\end{figure}
Examining the transmittance, it is clear that at 28 GHz, the power is maximized to the desired FB mode ($\text{T}_{-1}$), while the specular transmission ($\text{T}_{0}$) is suppressed. Additionally, the proposed metasurface is near-reflectionless at 28 GHz, with less than 2\% of the power contained in both reflected modes ($\text{R}_{0}$, $\text{R}_{-1}$). The electric field distribution portrayed in Figure \ref{fig11} at 28 GHz reveals that anomalous refraction is clearly achieved. 
\subsection{Total and Refraction Efficiencies}
\begin{figure}[t]
	\centering
	\subfloat[]{
		\includegraphics[scale=0.8]{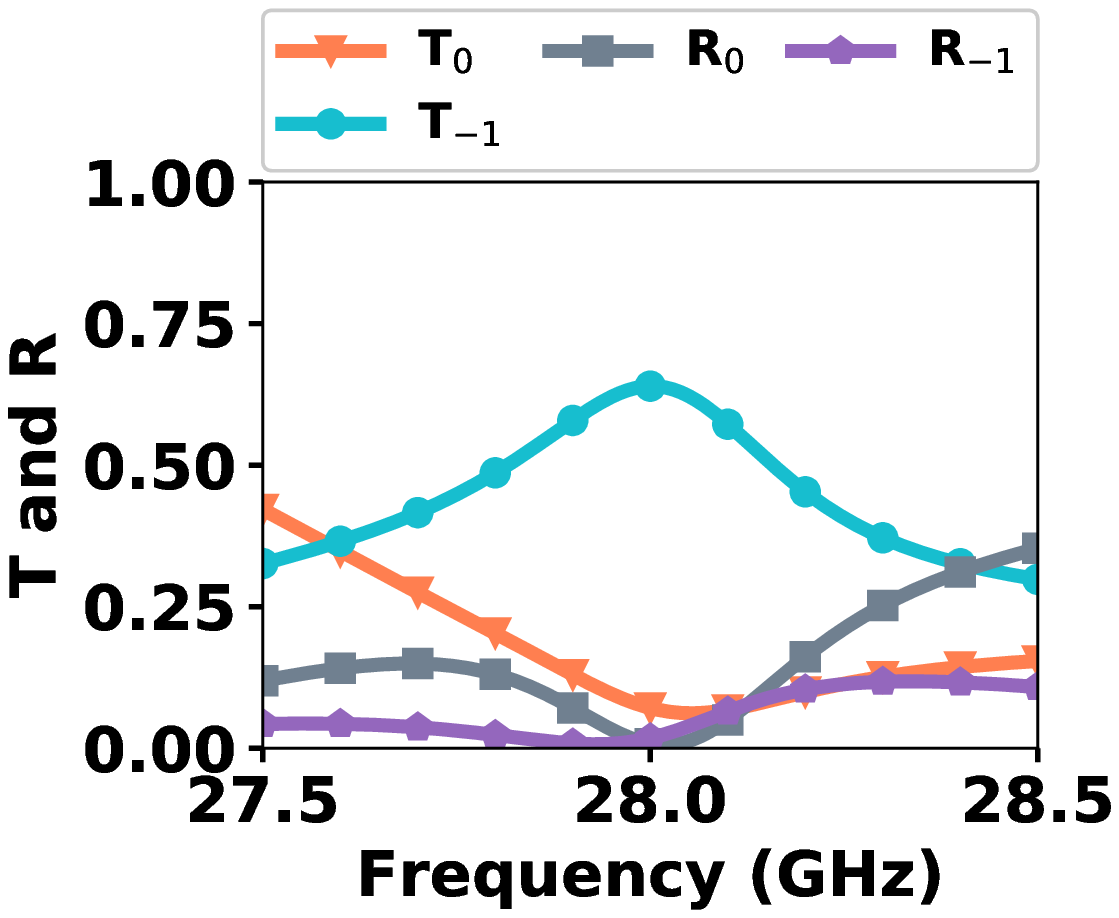}}
	\caption{\label{fig12} Transmittance and Reflectance of bipartite dielectric Huygens' metasurface with dielectric loss ($\tan\delta=0.003$).}
\end{figure}

So far, the metasurface has been designed without considering the dielectric losses. To determine the efficiency (total and refracted) of the metasurface, we have simulated the proposed Bi-DHMS with a typical dielectric loss tangent of 0.003~\cite{pre2021}. Figure \ref{fig12} shows the transmittance and reflectance plots for the case mentioned above.
The refraction efficiency and total efficiency can be computed as~\cite{mc2018}
\begin{equation}\label{eq7}
	\eta_R=\frac{P_{s(-1)}}{P_{ts}},
\end{equation}
and
\begin{equation}
	\eta_T=\frac{P_{ts}}{P_i},
\end{equation}
where $P_{ts}$ is the total scattered power, $P_i$ is the incident power, and $P_{s(-1)}$ represent the scattered power in the refracted beam. Using the above definitions, the calculation at 28 GHz reveals that 87\% of total scattered power is coupled to the desired $n=-1$ mode, which propagates towards $\theta_t=-44.5\degree$. In addition, the total efficiency of the metasurface is 75\%, showing that the dielectric absorbs 25\% of the incident power. The proposed metasurface shows good efficiency; however, utilizing a lower-loss dielectric material can further enhance the total efficiency.
\section{Conclusion}
In this paper, we have reported a \emph{bipartite dielectric Huygens' metasurface} -- a metasurface with two meta-atoms per period, capable of performing near-reflectionless anomalous refraction at 28 GHz. Through full-wave simulations, it has been shown that the proposed metasurface redirects the EM wave impinging at $\theta_i=15\degree$ to the desired FB mode, propagating towards $\theta_t=-44.5\degree$, with the refraction efficiency of 87\%. In comparison to previous proposals, the bipartite dielectric Huygens' metasurface introduced here is simple with a larger feature size that suits practical fabrication even at high frequencies, such as mm-wave and beyond, where constructing metasurfaces with small feature sizes is challenging. The coarse discretization paves the way to design simple yet efficient, cost-effective, and robust metasurfaces for the next-generation (5G \& beyond) wireless communication systems. In addition, the proposed metasurface is free from the conductive elements and can be scaled to other parts of the electromagnetic spectrum, such as terahertz and optical frequencies, enabling efficient and low-loss manipulation of light via all-dielectric nanophotonic devices.

\section*{Data availability statement}
The data that support the findings of this study are available upon reasonable request from the authors.

\section*{Conflict of interest}
The authors declare no conflict of interest.

\section*{Acknowledgements}
This work was supported by an Early Career Scheme from the Research Grants Council of the Hong Kong under Grant 21211619, and a Research Matching Grant Scheme under Grant 9229058.

\bibliographystyle{ieee}
\bibliography{ms}

\end{document}